\documentclass[aps,prd,twocolumn,showpacs,eqsecnum,A4,amsmath,amssymb, nofootinbib]{revtex4-1}

\usepackage{graphicx}

\usepackage{color}

\def \hs {\hspace{1.0mm}}

\newcommand{\Dif}{{\rm D}}
\newcommand{\dd}{{\rm{d}}}

\newcommand{\X}{{\rho}}
\def \H {\mathcal{H}}

\def \B {\mathcal{B}}
\def \pul {\textstyle{\frac{1}{2}}}

\def \be {\begin{equation}}
\def \ee {\end{equation}}
\def \bea {\begin{eqnarray}}
\def \eea {\end{eqnarray}}

\newcommand{\boldu}{\mbox{\boldmath$u$}} 
\newcommand{\bolde}{\mbox{\boldmath$e$}} 

\begin{document}

\title{Exact black holes in quadratic gravity with any cosmological constant}

\author{R.~\v{S}varc, J.~Podolsk\'y}
\email{robert.svarc@mff.cuni.cz}
\email{podolsky@mbox.troja.mff.cuni.cz}
\affiliation{
  Institute of Theoretical Physics, Charles University, Prague,
  Faculty of Mathematics and Physics, V~Hole\v{s}ovi\v{c}k\'ach~2, 180~00 Prague 8, Czech Republic}
\author{V.~Pravda, A.~Pravdov\' a}
\email{pravda@math.cas.cz}
\email{pravdova@math.cas.cz}
\affiliation{
  Institute of Mathematics of the Czech Academy of Sciences,
  \v Zitn\' a 25, 115 67 Prague 1, Czech Republic}

\date{\today}

\begin{abstract}

\noindent We present a new explicit class of black holes in 
general quadratic gravity with a
 cosmological constant. These spherically symmetric 
Schwarzschild-Bach-(anti--)de Sitter geometries (Schwa-Bach-(A)dS), 
derived under the assumption of constant scalar curvature,
form a three-parameter family determined by the black-hole horizon position, 
the value of Bach invariant on the horizon, and the cosmological constant. 
Using a conformal to Kundt metric ansatz, the fourth-order field equations 
simplify to a compact autonomous system. Its solutions are found as power series, 
enabling us to directly set the Bach parameter and/or cosmological constant equal 
to zero. To interpret these spacetimes, we analyse the metric functions. 
In particular, we demonstrate that for a certain range of positive cosmological 
constant there are both black-hole and cosmological horizons, with a static 
region between them. The tidal effects on free test particles and basic 
thermodynamic quantities are also determined.
\end{abstract}



\maketitle

\section{Introduction}

Black holes, regions with very strong gravity from which not
even light can escape, are one of the most fascinating theoretical
predictions of Einstein's general
relativity~\cite{Einstein1916}. The first exact solution to
this theory was almost immediately found by Schwarzschild
\cite{Schwarzschild1916}, describing a static spherically
symmetric spacetime. However, it took several decades to fully
understand its black-hole nature. This initiated the `golden
age' of black hole studies, epitomized by the discovery of
astrophysically more relevant Kerr rotating solution
\cite{Kerr}. Studies of various aspects of these `collapsed
objects', such as influence on matter and fields, the
no-hair conjecture, thermodynamic properties, or quantum
evaporation, followed soon. Moreover, a great observational
effort brought the direct evidence of their existence in our
universe when Cygnus X-1 source was identified as a black hole.
Also, now it seems that supermassive black holes reside in
nuclei of almost all galaxies. Mergers of black-hole binaries
have been recently detected as the first gravitational wave
signals. 	

Another remarkable interplay between Einstein's theory and
observed astronomical phenomena is a concept of the
cosmological constant. The famous $\Lambda$-term was introduced
by Einstein into his field equations to allow a static
cosmological model \cite{Einstein1917}. However, it was soon
demonstrated by de~Sitter \cite{deSitter1917} that the
cosmological constant causes even an empty space to expand
exponentially fast \cite{Schroedinger1956}. Nowadays, this is
employed for a phenomenological description of the observed
accelerated expansion of our universe caused by `dark energy'.
The de~Sitter solution also captures main features of the
inflationary epoch in the very early universe.

Despite all the great successes of Einstein's gravity theory,
it also has its limits, in particular, impossibility to
quantize it in the same way as other fundamental interactions,
and perhaps some open cosmological issues. Various extensions of
general relativity have thus been considered, see
\cite{Sotiriou:2010, DeFelice:2010, Capozziello:2011, Clifton:2012} for reviews.
In these modified theories, the black hole solutions play a
prominent role, providing natural test beds for their
comparison \cite{Tangherlini:1986, MP:1986, BoulwareDeser:1985, LuPerkinsPopeStelle:2015}.

Assuming a constant scalar curvature, we derive a \emph{new class} of
static spherically symmetric black hole solutions with a
cosmological constant $\Lambda$ in  quadratic
	gravity~\cite{Stelle:1978}, which includes the Einstein-Weyl theory~\cite{Weyl1919,Bach1921}. It generalizes
\cite{Kottler:1918} to include higher-order gravity
corrections, and \cite{LuPerkinsPopeStelle:2015,
	PodolskySvarcPravdaPravdova:2018a} to admit any $\Lambda$. 
		In contrast with the black holes of \cite{LuPerkinsPopeStelle:2015,
			PodolskySvarcPravdaPravdova:2018a}, the second (cosmological) horizon may appear due to ${\Lambda>0}$.
		 On large scales, the higher-order  corrections considerably affect the asymptotic behaviour of the geometry, which, even in the case of ${\Lambda=0}$, is not asymptotically flat (except for finely tuned parameters). This additional freedom thus opens completely new and more involved possibilities.  Moreover, both the cosmological constant and higher-order corrections are of key importance in quantum gravity models,  e.g.,~\cite{CoPe:2006}.

Within this setting, the vacuum action of quadratic gravity
contains $\Lambda$, the Ricci scalar $R$, and a contraction of the Weyl tensor $C_{abcd}$, namely
\be
S = \int \dd^4 x\, \sqrt{-g}\, \Big(\gamma \,
(R-2\Lambda) +\beta\,R^2  - \alpha\, C_{abcd}\, C^{abcd}\Big) \,, \label{actionQG}
\ee
where $\alpha$, $\beta$, ${\gamma=G^{-1}}$ are
constants. The corresponding field equations read
\begin{align}
&\gamma \left(R_{ab} - {\pul} R\, g_{ab}+\Lambda\,g_{ab}\right)-4 \alpha\,B_{ab} \nonumber \\
&\quad +2\beta\left(R_{ab}-\tfrac{1}{4}R\, g_{ab}+ g_{ab}\, \Box - \nabla_b \nabla_a\right) R = 0 \,, \label{GenQGFieldEq}
\end{align}
where
${B_{ab} \equiv \big( \nabla^c \nabla^d + {\pul} R^{cd} \big)C_{acbd}}$ 
is the traceless, symmetric and conserved~\emph{Bach~tensor}. 
Assuming ${R=\hbox{const.}}$, the last term in (\ref{GenQGFieldEq}) simplifies and the trace of the field
equations implies ${R=4\Lambda}$, so that they become
\be
R_{ab}-\Lambda\,g_{ab}=4k\, B_{ab}\,, \qquad \hbox{with}\qquad k \equiv \frac{\alpha}{\gamma+8\beta\Lambda} \,,
\label{fieldeqsEWmod}
\ee
see \cite{PravdaPravdovaPodolskySvarc:2017}.
For ${k=0}$ vacuum Einstein's
equations with a cosmological constant are
obtained. For ${\beta=0}$ we get Einstein-Weyl gravity. For ${\gamma+8\beta\Lambda= 0}$ the conformal Weyl theory is restored, in which the rotational curves of galaxies were studied \cite{MaKa:1989} within the spherically symmetric setting. Our solution, as an unifying model, may enable the analysis of relations between these theories in such astrophysical situations. 


\section{The geometry}\label{BH metric}

A spherically symmetric metric is usually written~as
\begin{equation}
\dd s^2 = -h(\bar r)\,\dd t^2+\frac{\dd \bar r^2}{f(\bar r)}+\bar r^2(\dd \theta^2+\sin^2\theta\,\dd \phi^2) \,.
\label{Einstein-WeylBH}
\end{equation}
However, in \cite{PodolskySvarcPravdaPravdova:2018a,
	PodolskySvarcPravdaPravdova:2018b} it was shown that for
investigation of such geometries in quadratic gravity, an
\emph{alternative form is more convenient}, \be \dd s^2 =
\Omega^2(r)\big[\,\dd \theta^2+\sin^2\theta\,\dd \phi^2 -2\,\dd
u\,\dd r+{\cal H}(r)\,\dd u^2 \,\big]\,. \label{BHmetric} \ee
This is related to (\ref{Einstein-WeylBH}) via
\begin{equation}
\bar{r} = \Omega(r)\,, \qquad t = u - {\textstyle\int}\, {\H(r)}^{-1}\dd r \,, \label{to static}
\end{equation}
and the metric functions $\Omega$, $\H$ give $f$, $h$ using
\be
h({\bar r}) = -\Omega^2\, \H \,, \qquad f({\bar r}) = -\left(\frac{\Omega'}{\Omega}\right)^2 \H \,, \label{rcehf}
\ee
(prime denotes the derivative with respect to $r$). The new metric \eqref{BHmetric} is
\emph{conformal} to a simple direct-product Kundt `seed', ${\dd
	s^2 =\Omega^2\,\dd s^2_{\hbox{\tiny Kundt}}}$, which is of the
algebraic type D, see \cite{Stephanietal:2003,
	GriffithsPodolsky:2009, PravdaPravdovaPodolskySvarc:2017}.

In the metric \eqref{BHmetric}, the \emph{Killing horizons}
corresponding to ${\partial_u=\partial_t}$ are located at
specific radii $r_h$ satisfying
\begin{equation}
\H \big|_{r=r_h}=0\,. \label{horizon}
\end{equation}
Of course, via \eqref{rcehf} this gives ${h({\bar
		r_h})=0=f({\bar r_h})}$. There is a time-scaling freedom
${t\to\sigma^{-1}\, t}$ of the metric \eqref{Einstein-WeylBH}
implying ${h\to \sigma^2\, h}$, which can be used, e.g., to
adjust appropriate value of ${h}$ at a chosen radius.

To uniquely characterize the geometries \eqref{BHmetric}, we need the Weyl and Bach \emph{scalar curvature invariants},
\begin{align}
C_{abcd}\, C^{abcd} &=  \tfrac{1}{3}\,\Omega^{-4}\,({\cal H}'' +2)^2 \,, \label{invC} \\
B_{ab}\, B^{ab} &=  \tfrac{1}{72}\,\Omega^{-8}\big[(\B_1)^2+2(\B_1+\B_2 )^2\big] \,,\label{invB}
\end{align}
where \emph{two independent Bach components} are
\be
\B_1 \equiv \H \H''''\,, \qquad \B_2 \equiv \H'\H'''-\tfrac{1}{2}{\H''}^2+2\,. \label{B2}
\ee
Interestingly, ${ B_{ab}=0 \ \Leftrightarrow \ B_{ab}\, 
	B^{ab}=0}$. Thus  we distinguish two geometrically
different types of solutions in quadratic gravity defined
by ${B_{ab}=0}$ and ${B_{ab}\ne0}$, respectively.

\section{The field equations}\label{derivingFE}

Under conformal transformations, the Bach tensor simply scales as
${B_{ab} = \Omega^{-2}\,B_{ab}^{\hbox{\tiny Kundt}}}$ and since higher-order corrections in \eqref{fieldeqsEWmod} are represented by the Bach tensor, using the metric \eqref{BHmetric} leads to a remarkable simplification of the field equations.
 Explicit
evaluation of the field equations \eqref{fieldeqsEWmod} for
\eqref{BHmetric}, using the Bianchi identities, yields two
simple ODEs for the metric functions $\Omega(r)$ and ${\cal
H}(r)$, namely
\begin{align}
\Omega\Omega''-2{\Omega'}^2 = &\ \tfrac{1}{3}k\, \B_1 \H^{-1} \,, \label{Eq1}\\
\Omega\Omega'{\cal H}'+3\Omega'^2{\cal H}+\Omega^2 -\Lambda\Omega^4 = &\ \tfrac{1}{3}k \,\B_2  \,, \label{Eq2}
\end{align}
see \cite{PPPS:2018d} for more details. It is also convenient to express the trace of \eqref{fieldeqsEWmod}, namely  ${R=4\Lambda}$,
\begin{equation}
{\cal H}\Omega''+{\cal H}'\Omega'+{\textstyle \frac{1}{6}} ({\cal H}''+2)\Omega = \tfrac{2}{3}\Lambda\,\Omega^3 \,. \label{trace}
\end{equation}
In fact, it is the derivative of \eqref{Eq2} minus ${\H'}$ times~\eqref{Eq1}.
The crucial point for further investigations is that
Eqs.~\eqref{Eq1}, \eqref{Eq2} do not explicitly depend
on~$r$. Solutions to such an \emph{autonomous system} can thus
be found as \emph{power series} in $r$ expanded around
\emph{any} point~${r_0}$
\be
\Omega(r) = \Delta^n \,
\sum_{i=0}^\infty a_i \,\Delta^{i}\,, \qquad \H(r) = \Delta^p
\, \sum_{i=0}^\infty c_i \,\Delta^{i}\,, \label{rozvoj}
\ee
where ${\Delta\equiv r-r_0}$, ${n,p\in \mathbb{R}}$, and ${a_0,\,c_0\neq0}$.

\subsection{Vanishing Bach tensor} \label{integration:Schw}

For ${\B_1=0=\B_2}$, we deal with Einstein's theory, and
Eqs.~\eqref{Eq1}, \eqref{Eq2} can be directly integrated. Using
the gauge freedom ${r \to \lambda\,r+\nu}$, ${u
	\to\lambda^{-1}u}$ of the metric \eqref{BHmetric}, this
immediately implies
\begin{equation}
\Omega(r)= \bar{r} = -\frac{1}{r}\,, \qquad \H(r) = \frac{\Lambda}{3}-r^2-2m\, r^3 \,, \label{SchwAdS}
\end{equation}
where the mass parameter $m$ is fixed by (\ref{horizon}), see (\ref{SchwAdSH}).
These functions represent the \emph{Schwarzschild-(anti--)de
	Sitter} spacetime \cite{Kottler:1918,Stephanietal:2003,
	GriffithsPodolsky:2009} which, expressed in the form
\eqref{Einstein-WeylBH} using \eqref{rcehf}, reads
${f=h=1-2m\,{\bar{r}}^{-1}-\frac{1}{3}\Lambda\,\bar{r}^2}$.

It is well known \cite{GriffithsPodolsky:2009} that for
${0<9\Lambda m^2<1}$ there are \emph{two horizons} determined
by \eqref{horizon}, namely the \emph{black-hole event horizon}
at $r_h$ and the \emph{cosmological horizon} at ${r_c>r_h}$
(they degenerate to ${{\bar r}_h={\bar r}_c=3m=1/\sqrt\Lambda}$
when ${9\Lambda m^2=1}$; ${\Lambda<0}$ admits only the black
hole horizon).

\subsection{Non-vanishing Bach tensor} \label{integration:nonSchw}

In a generic case (${\B_1, \B_2 \ne0}$), the system
\eqref{Eq1}, \eqref{Eq2} becomes non-trivially coupled but its
solutions can be found in the form \eqref{rozvoj}. Substituting
these series into the field equations, we obtain polynomial expressions where the dominant (lowest) powers of
$\Delta$ immediately put specific restrictions on the parameters
${[n,\,p]}$ and the possible value of $\Lambda$,
see Tab.~\ref{table:npLambda} and \cite{PPPS:2018d}.
In the next section, we will discuss the most
interesting case ${[0,\, 1]}$ corresponding to a \emph{single} root ${r_0}$ of (\ref{horizon}). 

\begin{table}[htb]
\begin{center}
\caption{\label{table:npLambda} The only admitted parameters ${[n,\, p]}$ in \eqref{rozvoj}, and the cosmological constant ${\Lambda}$, restricted by dominant powers of $\Delta$ in the field equations \eqref{Eq1}, \eqref{Eq2}, and the trace \eqref{trace}. Note that in the last column, $n\not=-1,-1/2$.}
\vspace{2.0mm}
\begin{tabular}{c|ccccccccc}
$n$\hs 			&\hs 0	&\hs 0	&\hs 1	&\hs $-1$	&\hs $-1$				&\hs 0											 &\hs 0 							&\hs	${<0}$ \\ \hline
$p$\hs			&\hs 1	&\hs 0	&\hs 0	&\hs  2	&\hs 0				&\hs 2				&\hs ${\ge 2}$				 							 &\hs ${2n+2}$ \\ \hline
$\Lambda$\hs&\hs any&\hs any&\hs any&\hs  0	&\hs ${\neq0}$&\hs ${\neq0}$&\hs ${\frac{3}{8k}}$&\hs $\frac{11n^2+6n+1}{1-4n^2}{\frac{3}{8k}}$\\
\end{tabular}
\end{center}
\end{table}

\section{Explicit black holes\label{ExplicitBH}}

In the case ${n=0}$, ${p=1}$, the root of~$\H$ representing the
non-degenerate Killing horizon \eqref{horizon} is explicitly
given by ${r_0\equiv r_h}$. The field equations (\ref{Eq1}),
(\ref{Eq2}), with (\ref{trace}), then restrict the coefficients
in the expansions \eqref{rozvoj} as
\begin{align}
&a_1 =\frac{1}{3c_0}\left[2\Lambda a_0^3-a_0(1+c_1)\right] , \nonumber \\
&c_2 =\frac{1}{6kc_0}\left[a_0^2(2-c_1-\Lambda a_0^2)+2k(c_1^2-1)\right] , \nonumber \\
&a_{l}=\frac{1}{l^2c_0}\Big[\,\tfrac{2}{3}\Lambda \sum^{l-1}_{j=0}\sum^{j}_{i=0}a_i\,a_{j-i}\,a_{l-1-j}-\tfrac{1}{3}\,a_{l-1} \nonumber\\
& \hspace{10.0mm} -\sum^{l}_{i=1}c_i\,a_{l-i}\left(l(l-i)+\tfrac{1}{6}i(i+1)\right)\Big] \,, \label{01coeff} \\
&c_{l+1}=\frac{3}{k(l+2)(l+1)l(l-1)} \nonumber \\
&\hspace{10.0mm} \times \sum^{l-1}_{i=0}a_i\, a_{l-i}(l-i)(l-1-3i) \,,\quad \hbox{for}\quad l\geq2\,,\nonumber
\end{align}
with three free parameters ${a_0,\  c_0,\  c_1}$.

To identify the Schwarzschild-(anti--)de Sitter spacetime
\eqref{SchwAdS} in the form (\ref{rozvoj}) with (\ref{01coeff}), first we  evaluate the Bach tensor \eqref{B2} on
the horizon, yielding ${\B_1(r_h) = 0}$, ${\B_2(r_h) =
	-\frac{3}{k}a_0^2\,b}$, where
${b\equiv\frac{1}{3}(c_1-2+\Lambda a_0^2)}$. Interestingly, by
setting ${b=0}$ (i.e. for ${c_1=2-\Lambda a_0^2}$), the Bach
tensor vanishes \emph{everywhere}. Employing the gauge freedom of
\eqref{BHmetric}, we may also set
\begin{equation}
a_0 =-\frac{1}{r_h} \,, \qquad c_0 =r_h-\frac{\Lambda}{r_h} \,. \label{01background}
\end{equation}
The explicit solution \eqref{rozvoj}, \eqref{01coeff} for
${b=0}$ then becomes
\begin{equation}
\Omega(r)=-\frac{1}{r}\,, \qquad \H(r) = \frac{\Lambda}{3}-r^2-\Big(\,\frac{\Lambda}{3}-r_h^2\Big)\,\frac{r^3}{r_h^3} \,, \label{SchwAdSH}
\end{equation}
where the expansions (\ref{rozvoj}) were summed-up as geometric series. This is exactly the \emph{Schwarzschild-(anti--)de Sitter black hole} \eqref{SchwAdS} since ${\frac{\Lambda}{3}-r_h^2=2m\,r_h^3}$.

In the case ${b\neq0}$, we may now separate the `Bach contribution'
in the coefficients \eqref{01coeff} proportional to $b$
 by introducing
$\alpha_i,\,\gamma_i$.
 With the same gauge choice \eqref{01background}, we
obtain a one-parameter \emph{extension} of the
Schwarzschild-(A)dS spacetime in quadratic gravity,
\begin{align}
\Omega(r) & = -\frac{1}{r}-\frac{b}{r_h}\sum_{i=1}^\infty\alpha_i\Big(\,\frac{r_h-r}{\X\,r_h}\Big)^i \,, \label{Omega_[0,1]}\\
\mathcal{H}(r) & = (r-r_h)\bigg[\,\frac{r^2}{r_h}-\frac{\Lambda}{3r_h^3}\left(r^2+rr_h+r_h^2\right) \nonumber \\
&\hspace{18.0mm} +3b\,\X\,r_h\sum_{i=1}^\infty\gamma_i\Big(\,\frac{r-r_h}{\X\,r_h}\Big)^i\,\bigg] \,, \label{H_[0,1]}
\end{align}
where
\begin{align}
\X &\equiv  1-\frac{\Lambda}{r_h^2}\,, \label{alphasgammainitial_[0,1]} \\
\alpha_1 &\equiv 1\,,\quad \gamma_1=1\,,\quad
\gamma_2 = \frac{1}{3}\Big[4-\frac{1}{r_h^2}\Big(2\Lambda+\frac{1}{2k}\Big)+3b\Big] \,, \nonumber
\end{align}
and $\alpha_l,\,\gamma_{l+1}$ are (with ${\alpha_0\equiv 0}$) \emph{recursively} given by
\begin{widetext}

\vspace{-6.0mm}

\begin{align}
&\alpha_{l}= \, \frac{1}{l^2}\Big[-\frac{2\Lambda}{3r_h^2}\,\sum_{j=0}^{l-1}\sum_{i=0}^{j}\big[\alpha_{l-1-j}\,\X^j+\big(\X^{l-1-j}+b\,\alpha_{l-1-j}\big)\big(\alpha_i\, \X^{j-i}+\alpha_{j-i}(\X^i+b\,\alpha_{i})\big)\big]-\tfrac{1}{3}\alpha_{l-2}(2+\X)\X(l-1)^2 \nonumber\\
& \hspace{14.0mm} +\alpha_{l-1}\big[\tfrac{1}{3}+(1+\X)\big(l(l-1)+\tfrac{1}{3}\big)\big]-3\sum_{i=1}^{l}(-1)^i\,\gamma_i\,(\X^{l-i}+b\,\alpha_{l-i})\big(l(l-i)+\tfrac{1}{6}i(i+1)\big)\Big]\,, \nonumber\\
&\gamma_{l+1}= \, \frac{(-1)^{l}}{kr_h^2\,(l+2)(l+1)l(l-1)}\sum_{i=0}^{l-1}\big[\alpha_i \,\X^{l-i}+\alpha_{l-i}\big(\X^i+b\,\alpha_i\big) \big](l-i)(l-1-3i) \,, \quad \hbox{for}\quad l\geq2\,. \label{alphasgammasgeneral_[0,1]}
\end{align}

\vspace{-6.0mm}

\end{widetext}

All these solutions form a \emph{three-parameter family of
spherically symmetric  black holes} (with static regions). In particular:
\begin{itemize}
\item The radius ${r=r_h}$ determines the \emph{Killing
    horizon} since ${\H(r_h)=0}$, see \eqref{H_[0,1]},
    \eqref{horizon}.
\item The  parameter ${\Lambda=R/4}$ is the \emph{cosmological constant}. It can be zero, recovering the results of \cite{PodolskySvarcPravdaPravdova:2018a}.
\item The \emph{Bach parameter} $b$ determines the Bach tensor
    contribution. For ${b=0}$, this
    Schwa-Bach-(A)dS black hole \eqref{Omega_[0,1]}, \eqref{H_[0,1]} reduces
    to \eqref{SchwAdSH}.
\end{itemize}

In terms of these three physical parameters, the scalar
    invariants \eqref{invC}, \eqref{invB} on the horizon
    are
\begin{align}
C_{abcd}\, C^{abcd}(r_h) &= 12\,\big((1+b)r_h^2-\tfrac{1}{3}\Lambda\big)^2 \,, \label{CinvBinv1}\\
B_{ab}\,B^{ab}(r_h) &= \frac{r_h^4}{4 k^2}\,b^2 \,. \label{CinvBinv2}
\end{align}

\begin{figure}[h!]
\includegraphics[scale=0.44]{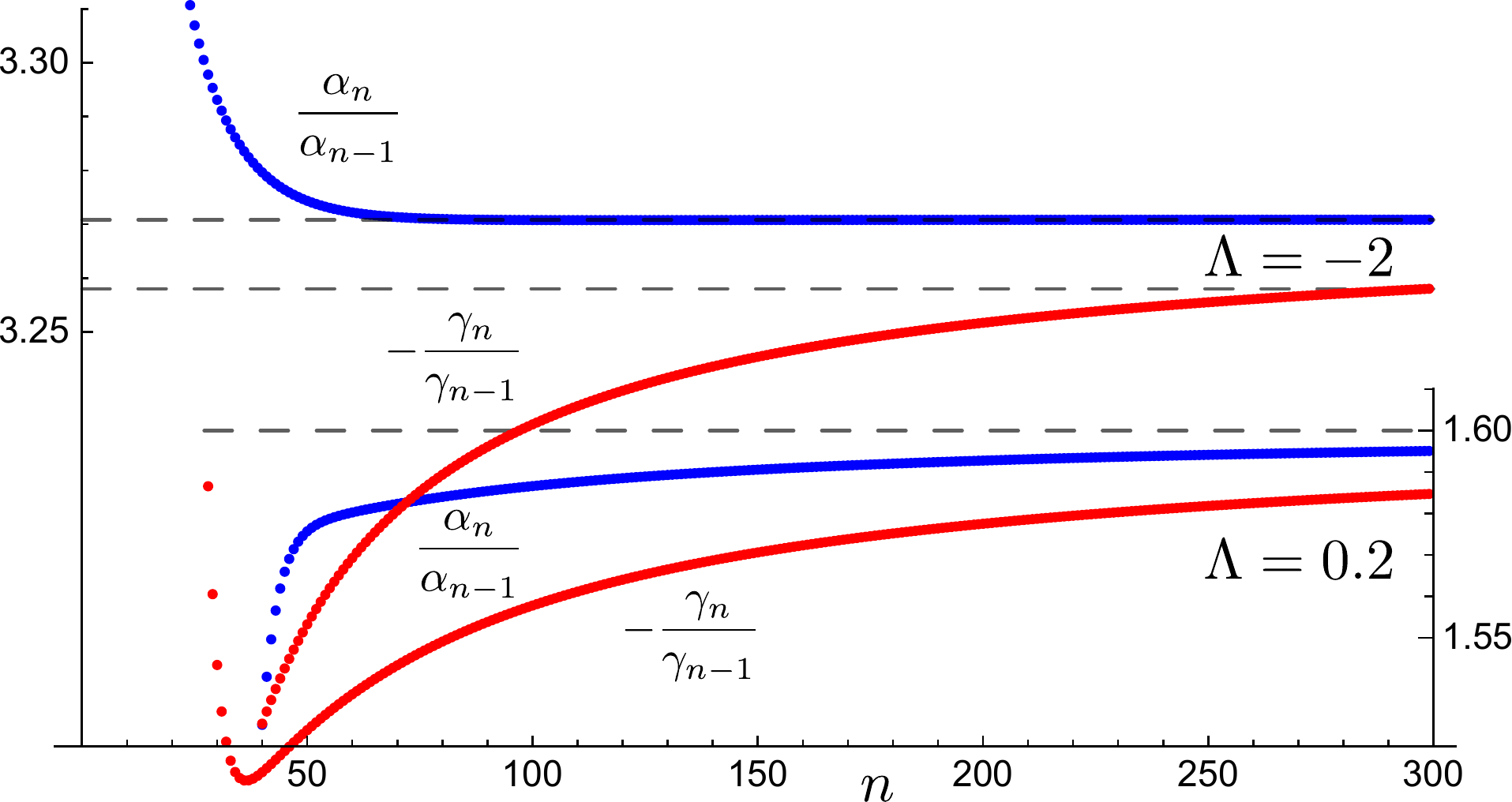}
\caption{\label{fig:1}
The convergence radius can be estimated from the ratio convergence test for solutions \eqref{Omega_[0,1]}, \eqref{H_[0,1]}, here given by ${r_h=-1,\, k=0.5}$ with ${b=0.3,\, \Lambda=0.2}$ (bottom) and ${b=0.2,\, \Lambda=-2}$ (top).}
\vspace{8mm}
\includegraphics[scale=0.44]{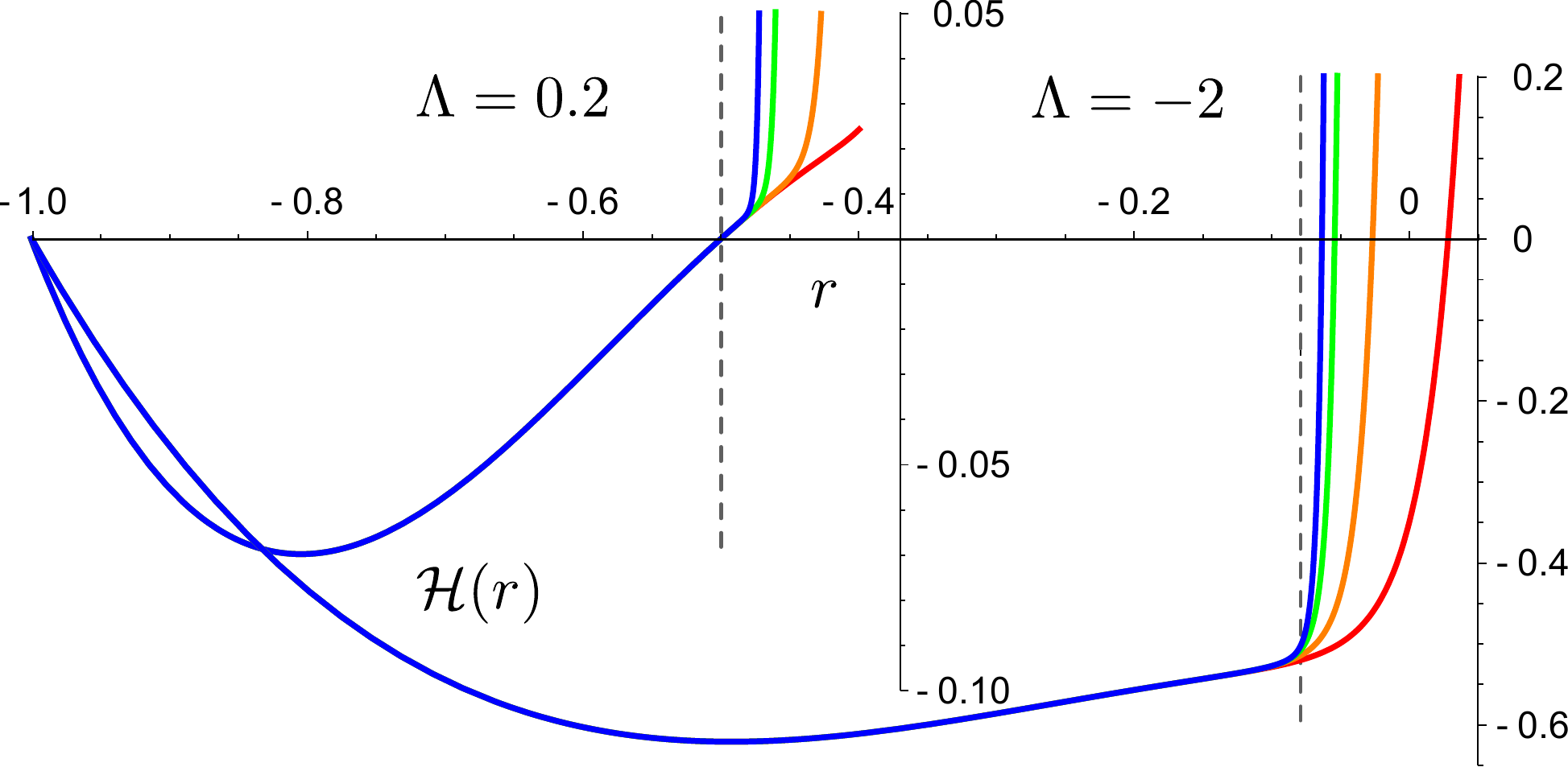}
\caption{\label{fig:2}
The function ${\cal H}(r)$ given by \eqref{H_[0,1]} for two values of the cosmological constant ${\Lambda}$ (with the same parameters as in Fig.~\ref{fig:1}). Both plots start on the black-hole horizon ${r_h=-1}$ and are reliable up to the vertical dashed lines indicating the radii of the convergence. For ${\Lambda>0}$ the function ${\cal H}(r)$ seems to have another root corresponding to the cosmological horizon, while for ${\Lambda<0}$ it remains non-vanishing. First 50 (red), 100 (orange), 200 (green), 300 (blue) terms in the expansions are used. The results fully agree with the numerical solutions up to the dashed lines, where such simulations also fail.}
\vspace{8mm}
\includegraphics[scale=0.44]{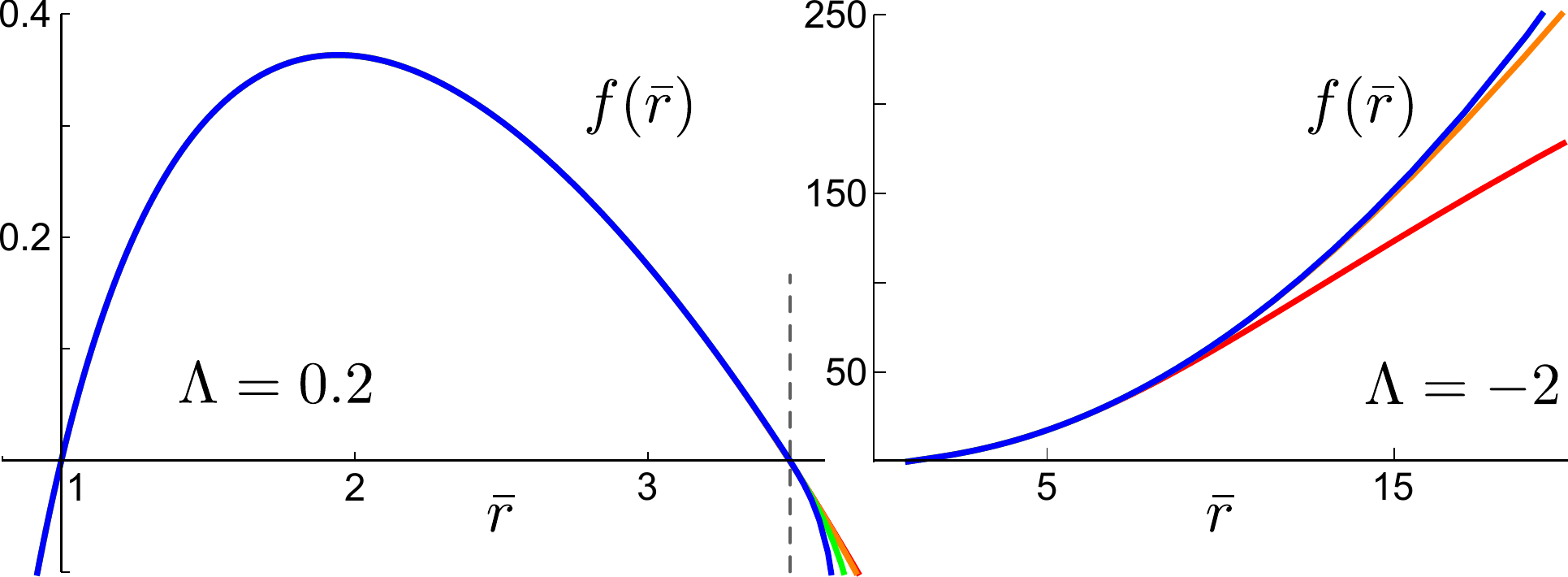}
\caption{\label{fig:3}
The function ${f(\bar{r})}$ of standard line element \eqref{Einstein-WeylBH} corresponding to the solution \eqref{Omega_[0,1]}, \eqref{H_[0,1]} via \eqref{rcehf} (with the same parameters as in Figs.~\ref{fig:1}, \ref{fig:2}). The ${\Lambda>0}$ case (left) indicates the presence of the cosmological horizon at the boundary of the convergence interval (the dashed line). For ${\Lambda<0}$ (right), the series converge in the whole plotted range, indicating a static region everywhere above the black-hole horizon.}
\end{figure}

In Fig.~\ref{fig:1}, convergence of the series in
\eqref{Omega_[0,1]}, \eqref{H_[0,1]} is examined using the
d'Alembert ratio test for two different sets of parameters. It
clearly indicates that, with $n$ growing, the ratio between two
subsequent terms approaches a specific constant. The series
thus \emph{asymptotically behave as geometric series}. This
enables us to estimate the radius of convergence.

Typical behaviour of the metric function ${\cal H}(r)$ outside
the black-hole horizon is plotted in Fig.~\ref{fig:2}. There is
a significant qualitative difference between ${\Lambda<0}$ and
${\Lambda>0}$. In both cases, the black-hole horizon separates
static (${r>r_h}$) and non-static (${r<r_h}$) regions of the
spacetime. However, for ${\Lambda>0}$ an outer boundary of this
static region appears, which corresponds to the cosmological
horizon given by the second root of $\H$ (as in the classic
Schwarzschild-de Sitter black hole). This is also demonstrated
in Fig.~\ref{fig:3} by plotting the function ${f(\bar{r})}$ of
the common metric \eqref{Einstein-WeylBH}.

\section{Specific tidal effects\label{GeodDev}}

The two independent parts \eqref{B2} of the Bach tensor $\B_1,
\B_2$ can be observed via a \emph{specific relative
	motion of free test particles} described by the equation of
geodesic deviation \cite{PodolskySvarc:2012}. For an invariant
description, we employ an orthonormal frame associated with
\emph{initially static observer}
(${\dot{r}=\dot{\theta}=\dot{\phi}=0}$) with velocity
${\boldu=\dot{u}\,\partial_u\equiv\bolde_{(0)}}$, namely
${\bolde_{(1)}=-\dot{u}\,(\partial_u+\H\,\partial_r)}$,
${\bolde_{(2)}=\Omega^{-1}\partial_\theta}$, and
${\bolde_{(3)}=(\Omega\sin\theta)^{-1}\partial_\phi}$.
Projection of the equation of geodesic deviation onto this
frame gives
\begin{align}
\ddot{Z}^{(1)} = &\, \frac{\Lambda}{3}\,Z^{(1)}+\frac{1}{6} \frac{{\cal H}''+2}{\Omega^2}\,Z^{(1)}-\frac{k}{3}\,\frac{\B_1+\B_2}{\Omega^4}\,Z^{(1)} , \label{InvGeoDevBH1r0}\\
\ddot{Z}^{(i)} = &\, \frac{\Lambda}{3}\,Z^{(i)}-\frac{1}{12}\frac{{\cal H}''+2}{\Omega^2}\,Z^{(i)}-\frac{k}{6}\,\frac{\B_1}{\Omega^4}\,\,Z^{(i)} , \label{InvGeoDevBHir0}
\end{align}
where ${i=2,3}$, ${Z^{(a)} \equiv {e^{(a)}}_{\!\!\mu}\,Z^\mu}$
denotes \emph{relative position} of two particles, and ${\ddot
	Z^{(a)} \equiv {e^{(a)}}_{\!\!\mu}\,\frac{\Dif^2 Z^\mu}{\dd\,
		\tau^2}}$ their \emph{mutual acceleration}. In
\eqref{InvGeoDevBH1r0}, \eqref{InvGeoDevBHir0}, we easily
identify classic parts corresponding to the \emph{isotropic
	influence} of the cosmological constant $\Lambda$ and the
\emph{Newtonian tidal effect} caused by the \emph{Weyl tensor}
proportional to the square root of \eqref{invC}. Moreover, the
theory satisfying (\ref{fieldeqsEWmod}) admits \emph{two additional effects}
encoded in the non-trivial \emph{Bach tensor} components $\B_1,
\B_2$. The first of them affects particles in the
\emph{transverse} directions~$\partial_\theta,
\partial_\phi$, see \eqref{InvGeoDevBHir0}, while the second
one induces their \emph{radial} acceleration along
$\partial_{\bar r}$ via \eqref{InvGeoDevBH1r0}. Since
${\B_1(r_h)=0}$, \emph{on any horizon} there is \emph{only the
radial} effect caused by ${\B_2(r_h)}$.

\section{Thermodynamic quantities: horizon area, temperature, entropy\label{Thermodynamic}}

Let us also determine main thermodynamic properties of this
explicit family of spherically symmetric Schwa-Bach-(A)dS black
holes. The horizon is generated by the (rescaled) null Killing
vector ${\ell\equiv\sigma\partial_u=\sigma\partial_t}$ and thus
is located at ${r=r_h}$ where ${\H=0}$, cf.~\eqref{horizon},
\eqref{H_[0,1]}. Its \emph{area} is, using \eqref{BHmetric},
\eqref{Omega_[0,1]},
\be
{\cal A}  =4\pi\,r_h^{\,-2}= 4\pi\,{\bar r}_h^2\,, \label{horizon_area}
\ee
while its
\emph{surface gravity}
(${\kappa^2\equiv-\frac{1}{2}\,\ell_{\mu;\nu}\,\ell^{\,\mu;\nu}}$)
reads
\be
\kappa/\sigma = -\tfrac{1}{2}\,\H'(r_h) = -\tfrac{1}{2}\rho\, r_h
	=\tfrac{1}{2}\, {\bar r}_h^{\,-1}(1-\Lambda {\bar r}_h^2) \,.
\label{surface_gravity}
\ee
It is \emph{the same expression as in the Schwarzschild-(A)dS case}, independent of the Bach
parameter $b$.
The black-hole horizon \emph{temperature} ${T = \tfrac{1}{2\pi}\,\kappa}$ is thus
\be
T/\sigma = -\tfrac{1}{4\pi}\rho\,r_h
	= \tfrac{1}{4\pi}\,{\bar r}_h^{\,-1}(1-\Lambda {\bar
r}_h^2) \,.
\label{temperature}
\ee
%
This is zero for ${{\bar r}_h=1/\sqrt\Lambda}$ corresponding to the case of extreme Schwarzschild-de Sitter black hole for which the black-hole and cosmological horizons coincide at ${{\bar r}_h={\bar r}_c}$.

However, in higher-derivative theories, we must apply the
generalized definition of \emph{entropy} ${S=(2\pi/\kappa)\oint
\mathbf{Q}\,}$,  see \cite{Wald:1993}, where the Noether charge
2-form on the horizon~is
\begin{align}
&\mathbf{Q} = -\frac{\Omega^2\, \H'}{16\pi}\left[\gamma+\frac{4}{3}\Lambda(\alpha+6\beta)+\frac{4}{3}k\alpha\,\frac{\B_1+\B_2}{\Omega^4}\right]\!\!\bigg|_{r=r_h} \nonumber \\
& \hspace{50.0mm}\times\sin\theta\,\dd\theta\wedge\dd\phi \,. \label{Noether}
\end{align}
Evaluating the integral, using \eqref{horizon_area},
\eqref{surface_gravity}, \eqref{CinvBinv2} and ${r_h=-1/{\bar r}_h}$, we get
{\be
				S = \frac{1}{4}{\cal
			A}\left[\gamma+\frac{4}{3}\Lambda\,(\alpha+6\beta) -4\alpha\,\frac{b}{{\bar r}_h^{\,2}}\right]
		\,. \label{entropy} \ee
}

For the Schwarzschild black hole (${b=0,\, \Lambda=0}$) or in
the Einstein theory {(${\alpha=0,\ \beta=0}$)}, this reduces to the standard
expression ${S = \tfrac{1}{4G}\,{\cal A}}$. For ${\Lambda=0}$, the
results of
\cite{LuPerkinsPopeStelle:2015,PodolskySvarcPravdaPravdova:2018a}
are recovered. For the Schwarzschild-(A)dS black hole (${b=0}$)
in Einstein-Weyl gravity {(${\beta=0}$)}, we obtain ${S =
	\tfrac{1}{4G}\,{\cal A}\,\big(1+\tfrac{4}{3}k\Lambda \big)}$,
which agrees with the results of \cite{LuPope:2011}. In
critical gravity, defined by {${\beta=0}$}, ${\alpha=k\gamma}$, ${\Lambda=-\frac{3}{4k}<0}$, the
entropy is zero. Our formula \eqref{entropy} for entropy
generalizes all these expressions to the case of
Schwarzschild-Bach-(anti--)de~Sitter black holes when the Bach
tensor is non-vanishing, parameterized by ${b\ne0}$. In this
case, the \emph{entropy is non-zero even in critical gravity}.
For smaller black holes, the deviations from {${S =
\tfrac{1}{4}{\cal A}\,\big[\gamma+\tfrac{4}{3}\Lambda (\alpha+6\beta)\big]}$} are larger.

By replacing the root $r_h$ by $r_c$ in \eqref{Omega_[0,1]},
\eqref{H_[0,1]}, the solution is expanded around the
\emph{cosmological horizon}. Its temperature and entropy are
thus given by \eqref{temperature} and \eqref{entropy},
respectively, in which ${\bar r}_h$ is simply replaced by
${\bar r}_c$.

\vspace{-4.0mm}

\section{Acknowledgements}

This work has been supported by the Czech Science Foundation
Grant No. GA\v{C}R 17-01625S and the Czech-Austrian MOBILITY
grant 8J18AT02 (JP, R\v{S}), and the Research Plan RVO:
67985840 (VP, AP). We thank H.~Maeda for reading the manuscript. We are also grateful to anonymous referees for their very useful comments, observations and suggestions.


\begin{thebibliography}{10}


\bibitem{Einstein1916} Einstein~A 1916
    Die Grundlage der allgemeinen Relativit\"{a}tstheorie,
    {\em Ann. der Physik} {\bf 49} 769

\bibitem{Schwarzschild1916} Schwarzschild~K 1916
    \"Uber das Gravitationsfeld eines Massenpunktes nach der Einsteinschen Theorie,
    {\em Sitz. Preuss. Akad. Wiss. Berlin} {\bf 7} 189

\bibitem{Kerr} Kerr R P 1963 Gravitational field of a spinning mass as an example of algebraically special metrics,
		{\em Phys. Rev. Lett.} {\bf 11} 237

\bibitem{Einstein1917} Einstein~A 1917 Kosmologische Betrachtungen zur allgemeinen Relativit\"{a}tstheorie,
		{\em Sitz. Preuss. Akad. Wiss. Berlin} 142

\bibitem{deSitter1917} de~Sitter~W 1917 Over de relativiteit
    der traagheid: Beschouingen naar aanleiding van Einstein's hypothese, {\em Koninklijke Akademie van
    Wetenschappen te Amsterdam} {\bf 25} 1268; 1918 {\em Proc.~Akad.~Amsterdam} {\bf 19} 1217

\bibitem{Schroedinger1956} Schr\"odinger~E 1956
        {\em Expanding universes} (Cambridge: Cambridge University Press)

\bibitem{Sotiriou:2010} Sotiriou T P and Faraoni V 2010 $f(R)$
    theories of gravity, \emph{Rev. Mod. Phys.} \textbf{82} 451

\bibitem{DeFelice:2010} De Felice A and Tsujikawa S 2010 $f(R)$
    Theories, \emph{Living Rev. Relativity} \textbf{13} 3

\bibitem{Capozziello:2011} Capozziello S and De Laurentis M
    2011 Extended Theories of Gravity, \emph{Physics Reports} \textbf{509} 167

\bibitem{Clifton:2012}
Clifton T et al. 2012 Modified gravity and cosmology, \emph{Physics Reports} \textbf{513} 1

\bibitem{Tangherlini:1986} Tangherlini F R 1963 Schwarzschild Field in $n$ Dimensions and the Dimensionality of Space Problem,
		{\em Nuovo Cim.} {\bf 77} 636

\bibitem{MP:1986} Myers R C and Perry M J 1986 Black Holes in Higher Dimensional SpaceTimes, {\em Annals Phys.} {\bf 172} 304

\bibitem{BoulwareDeser:1985} Boulware D G and Deser S 1985 String-Generated Gravity Models,
		{\em Phys. Rev. Lett.} {\bf 55} 2656

\bibitem{LuPerkinsPopeStelle:2015} L\"u~H, Perkins~A, Pope~C~N
    and Stelle~K~S 2015 Black holes in higher derivative
    gravity, {\em Phys. Rev. Lett.} {\bf 114} 171601

\bibitem{Stelle:1978} Stelle~K~S 1978 Classical gravity with
    higher derivatives, {\em Gen. Relativ. Gravit.} {\bf 9} 353
		
\bibitem{Weyl1919} Weyl~H 1919
    Eine neue Erweiterung der Relativit\"{a}tstheorie,
    {\em Ann. der Physik} {\bf 59} 101

\bibitem{Bach1921} Bach~R 1921
    Zur Weylschen Relativit\"{a}tstheorie und der Weylschen Erweiterung des Kr\"{u}mmungstensorbegriffs,
    {\em Math. Zeitschrift} {\bf 9} 110
		
\bibitem{Kottler:1918} Kottler F 1918 \"Uber die physikalischen Grundlagen der Einsteinschen Gravitationstheorie, {\em Ann. Physik} {\bf 56} 401		

\bibitem{PodolskySvarcPravdaPravdova:2018a} Podolsk\'{y}~J,
    \v{S}varc~R, Pravda~V and Pravdov\'a~A 2018 Explicit black hole solutions in higher-derivative gravity,
		 {\em Phys. Rev. D} {\bf 98} 021502(R)
		
\bibitem{CoPe:2006} Codello A and Percacci R 2006 Fixed Points of Higher-Derivative Gravity,
		{\em Phys. Rev. Lett.} {\bf 97} 221301
		
\bibitem{PravdaPravdovaPodolskySvarc:2017} Pravda~V,
    Pravdov\'a~A, Podolsk\'{y}~J and \v{S}varc~R 2017 Exact
    solutions to quadratic gravity, {\em Phys. Rev. D} {\bf 95} 084025

	
\bibitem{PodolskySvarcPravdaPravdova:2018b} Podolsk\'{y}~J,
    \v{S}varc~R, Pravda~V and Pravdov\'a~A 2018
    (in preparation)
		
\bibitem{MaKa:1989} Mannheim P D and Kazanas D 1989 Exact vacuum solution to conformal Weyl gravity and galactic rotation curves,
		{\em Astrophysical Journal} {\bf 342} 635 

\bibitem{Stephanietal:2003} Stephani~H, Kramer~D,
    MacCallum~M, Hoenselaers~C, and Herlt~E 2003 {\em Exact
    Solutions of Einstein's Field Equations} (Cambridge: Cambridge University Press).

\bibitem{GriffithsPodolsky:2009} Griffiths~J and Podolsk\'{y}~J
    2009 {\em Exact Space-Times in Einstein's General Relativity} (Cambridge: Cambridge University Press)


\bibitem{PPPS:2018d} Pravda~V, Pravdov\'a~A, Podolsk\'{y}~J and \v{S}varc~R 2018
    (in preparation)

\bibitem{PodolskySvarc:2012} Podolsk\'{y}~J and \v{S}varc~R
    2012 Interpreting spacetimes of any dimension using
    geodesic deviation, {\em Phys. Rev. D} {\bf 85} 044057

\bibitem{Wald:1993} Wald~R~M 1993 Black hole entropy is the Noether charge,
		{\em Phys. Rev. D} {\bf 48} R3427(R)

\bibitem{LuPope:2011} L\"u~H and Pope~C~N 2011 Critical gravity in four dimensions,
    {\em Phys. Rev. Lett.} {\bf 106} 181302

\end{thebibliography}
\end{document}